\documentclass[aoas,MSNbibl,nameyear,dvips]{arximspdf}
\usepackage{graphicx}

\doi{10.1214/11-AOAS505}
\volume{6}
\issue{1}
\pubyear{2012}
\firstpage{1}
\lastpage{26}

\makeatletter

\def\bsuffix #1{#1}

\newcommand{\bl}{\mathrm{bl}}

\makeatother

\begin{document}
\begin{frontmatter}

\title{Estimating within-school contact networks to understand
influenza transmission\thanksref{T1}}
\runtitle{Estimating within-school contact networks}

\thankstext{T1}{Supported by the NIH/NIGMS MIDAS Grant U01-GM070749.}

\begin{aug}
\author[A]{\fnms{Gail E.} \snm{Potter}\corref{}\ead[label=e1]{gpotter@fhcrc.org}},
\author[B]{\fnms{Mark S.} \snm{Handcock}\ead[label=e2]{handcock@ucla.edu}},
\author[C]{\fnms{Ira~M.}~\snm{Longini,~Jr.}\ead[label=e3]{ilongini@ufl.edu}}
\and
\author[A]{\fnms{M. Elizabeth} \snm{Halloran}\ead[label=e4]{betz@fhcrc.org}}
\runauthor{Potter, Handcock, Longini, Jr. and Halloran}
\affiliation{Fred Hutchinson Cancer Research Center, University of
California, Los~Angeles and University of Washington,
Emerging Pathogens Institute, University of
Florida and University of Washington,
and Fred Hutchinson Cancer Research~Center and University of Washington}
\address[A]{G. E. Potter\\
M. E. Halloran\\
Fred Hutchinson Cancer Research Center\\
1100 Fairview Ave N, M2-C200\\
Seattle, Washington 98109-1024\\
USA\\
\printead{e1}\\
\hphantom{E-mail: }\printead*{e4}}
\address[B]{M. S. Handcock\\
Department of Statistics\\
University of California\\
8125 Math Sciences Bldg.\\
Box 951554\\
Los Angeles, California 90095-1554\\
USA\\
\printead{e2}}
\address[C]{I. M. Longini, Jr.\\
Emerging Pathogens Institute\\
University of Florida\\
P.O. Box 100009\\
Gainesville, Florida 32610-0009\\
USA\\
\printead{e3}} %adresu isvedimo komanda gale!
\end{aug}

% HISTORY:
\received{\smonth{7} \syear{2011}}
\revised{\smonth{8} \syear{2011}}

% ABSTRACT
%
\begin{abstract}
Many epidemic models approximate social contact behavior by assuming
random mixing within mixing groups (e.g., homes, schools and
workplaces). The effect of more realistic social network structure on
estimates of epidemic parameters is an open area of exploration. We
develop a detailed statistical model to estimate the social contact
network within a high school using friendship network data and a survey
of contact behavior. Our contact network model includes classroom
structure, longer durations of contacts to friends than nonfriends and
more frequent contacts with friends, based on reports in the contact
survey. We performed simulation studies to explore which network
structures are relevant to influenza transmission. These studies yield
two key findings. First, we found that the friendship network
structure important to the transmission process can be adequately
represented by a dyad-independent exponential random graph model
(ERGM). This means that individual-level sampled data is sufficient to
characterize the entire friendship network. Second, we found that
contact behavior was adequately represented by a static rather than
dynamic contact network. We then compare a targeted antiviral
prophylaxis intervention strategy and a grade closure intervention
strategy under random mixing and network-based mixing. We find that
random mixing overestimates the effect of targeted antiviral
prophylaxis on the probability of an epidemic when the probability of
transmission in 10 minutes of contact is less than 0.004 and
underestimates it when this transmission probability is greater than
0.004. We found the same pattern for the final size of an epidemic,
with a threshold transmission probability of 0.005. We also find
random mixing overestimates the effect of a grade closure intervention
on the probability of an epidemic and final size for all transmission
probabilities. Our findings have implications for policy
recommendations based on models assuming random mixing, and can inform
further development of network-based models.
\end{abstract}

% KEYWORDS
%
\begin{keyword}
\kwd{Contact network}
\kwd{epidemic model}
\kwd{influenza}
\kwd{simulation model}
\kwd{social network}.
\end{keyword}

\end{frontmatter}

%s1 ###
\section{Introduction}
Schools play an important role in transmission of infectious diseases, so
understanding the transmission process within schools can improve our
ability to plan effective interventions. School closure is known to
reduce disease transmission, as demonstrated by \citet{dennis},
\citet
{rodriguez} and \citet{hensschool}, but this approach is costly on both
an individual and societal level. Mathematical models show that
vaccinating school-aged children is an effective strategy when vaccine
supplies are limited; see, for example, \citet{loeb} and
\citet
{bastaschool}. %also yang and longinischool
When a new strain of influenza virus or other pathogen has emerged,
large-scale agent-based epidemic simulation models have been used to
predict epidemic spread and compare intervention strategies. The
methodology underlying these models is described in \citet
{halloran}, \citet{germann}, \citet{eubank1} and \citet
{ferguson}. These
models simulate human contact behavior, and disease may be transmitted
when an infectious person contacts a susceptible person. In most such
models, social contact behavior is approximated by random mixing within
classrooms and
schools, as well as homes, workplaces and other mixing groups. That is,
people contact other mixing group members with equal probability during each
time step. This process is a simplification of the true underlying
social structure.

Simulation studies have shown that network structure can influence
epidemic dynamics. Several papers have demonstrated the varying
influence of
clustering and repetition in contacts on disease spread for a range of
parameter values. Among these, \citet{eames}, \citet
{smieszek} and \citet
{duerr} simulate idealized,
simplified networks that are not informed by data on contact behavior. For
example, the number of contacts in their models is equal for all
individuals. \citet{miller} explores these network structures using
Episims, a realistic agent-based network
simulation model built from transportation, location, activity and
demographic data, but not directly informed by contact surveys
[\citet{eubank1}].
\citet{Keeling05jrsi} and \citet{read} explored the
influence of degree
distribution on
disease spread, where the degree of a person is the number of contacts
he/she makes. The former of these uses a contact survey of 49
respondents, while the latter performs simulations
based on idealized networks. The development of statistical techniques
to infer
detailed and realistically complex network models for face-to-face contacts
based on available survey data is a relatively new area. Recent work
with the multicountry European POLYMOD study, a diary-based survey of
contact behavior, has
inferred within-household contact networks [\citet{potter}] and
age-based mixing
matrices [\citet{mossong}, \citet{hens2009}], but we do not yet have a clear
picture of the entire
contact network, nor a complete understanding of the relevant network
structures for epidemic transmission.

Some papers have focused on characterizing within-school contact
behavior in the context of understanding disease transmission.
\citet{glass}
administered contact surveys in an American elementary, middle and high
school, and characterized contact duration and intensity by grade and
location. \citet{conlan} developed a new method to collect contact
network data and analyzed mixing patterns, clustering and other network
properties in 11 British primary schools. Although these studies
provide important information regarding contact behavior within
schools, neither develops a method for inference of the entire
within-school contact network. \citet{cauchemez} analyzed network and
symptom status data in a~fourth grade class during the H1N1 influenza
pandemic. They found that selective mixing by gender influences the disease
dynamics, but found no evidence for a playmate network or classroom neighbor
effect on the transmission probability. However, because the sample
size was small and asymptomatic and unobserved cases were not accounted for
in the analysis, their findings are not definitive.
\citet{Sthetal} describe a face-toface contact network in a
primary school using proximity sensor data. \citet{Saletal10} analyze
wireless sensor data to describe the contact network in an American
high school and demonstrate through simulation studies that using
network data to inform interventions can reduce the disease
burden.
\citet{eubank} demonstrate that modeling network structure
within schools in a
large-scale simulation model can impact global epidemic dynamics.

In this paper we develop a statistical model of a within-school contact network
in order to understand how social network structures within schools
influence disease transmission. In Section \ref{sectiondata} we
describe our two data sources: friendship network data from a high
school and a survey on contact behavior in high schools. Section \ref
{sectionmethods} describes our methodology to model the contact network
and compare epidemics based on this contact network to those under
random mixing. In Section \ref{subsectionoutline} we outline our method to
model the contact network conditional on the friendship network. In Section
\ref{subsectiondegrees} we describe how we estimate the contact degree
distribution from the contact survey, and in Section \ref{subsectionlink} we
describe how we model the contact network conditional on the degree
distribution. In Section \ref{subsectioncontactsimulation} we describe
how we simulate contact networks from our model, and we describe our
influenza simulation procedure in Section \ref{subsectionsimulation}. We then
compare performance of different variations of the contact network
model in Section \ref{subsectionselection}. In Section \ref{subsectionergm} we
present
our model for the friendship network itself. In Section \ref{subsectioncompare}
we describe our procedure to compare epidemics based on our network
model to random mixing under three different scenarios: no
intervention, a targeted antiviral prophylaxis intervention, and a
grade closure strategy. Results from these comparisons are presented in
Section \ref{sectionresults} and discussed in Section \ref{sectiondiscussion}.

%s2 ###
\section{Data}
\label{sectiondata} 
\begin{figure}[b]

\includegraphics{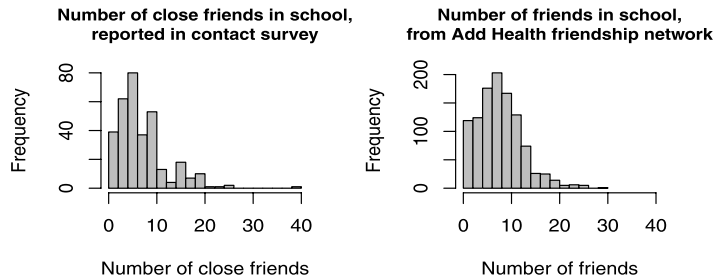}

\caption{Distribution of number of friends in the epidemic survey
(left) and in school 18 in the Add Health data set (right). The
different definitions of ``close friendship'' in the two data sources
produce similar distributions of number of close friends.}
\label{figdegreedist}
\end{figure}

We use two data sources to inform our contact
network model. The first is friendship network data from the Add Health
study, a survey of health, demographic and relational data administered
in 80 American high schools spanning grades 7--12, or high school plus
feeder school combinations for high schools not spanning those grades
[\citet{addhealthdata}]. The second was \textit{A Survey on
Epidemics in High Schools}, administered in two Virginia high schools
by the Network Dynamics and Simulation Science Laboratory at Virginia
Polytechnic Institute and State University during the spring of 2009
[\citet{eubank}]. The goal of the Add Health study was to survey
all students in each school [\citet{addhealthdesign}]. Prior to
the survey, each school created a school roster listing all students
with identification numbers. Students were given a copy of the roster
and identified their five best male friends and five best female
friends. Students could nominate friends not on the roster, and could
nominate fewer than five friends of each sex. In this paper, we analyze
one school configuration with 1,314 students. We selected this school
because it is fairly large and has less missing data than other
schools. We model contact behavior among the 1,074 students who
responded to the survey, were on the school roster on the survey date,
and have nonmissing grade values. We assume that two students are
friends if a~reciprocated or unreciprocated nomination occurred. By
defining friendship in this way, the friend degree distribution in this
data set is similar to that found in the contact survey. The two degree
distributions are compared in Figure \ref{figdegreedist}.

%f1 ###
%

Our contact data source, \textit{A Survey on Epidemics in High
Schools}, was~ad\-ministered in two Virginia high schools. In one, classes were
randomly sampled and the survey given to all consenting students in
the
sampled classes, resulting in a sample of 116 of 1,116 students. In the
other, the goal was to survey all 425 students, but only 246 students
returned the survey because interviewers did not explicitly state that
students were supposed to return the form. We'll refer to this from
here on as the ``epidemic survey.'' The survey defines a ``contact'' to
mean ``being in close
proximity for more than roughly five minutes.'' Respondents reported the
average number of contacts they make during class breaks and the lunch
break, the number of close friends they have in their school, and whether
students sitting near them in class are mostly close friends,
classmates but
not close friends or a~mix of the two. They also estimated the percentage
of contacts they made to friends.

\begin{figure}%[b]

\includegraphics{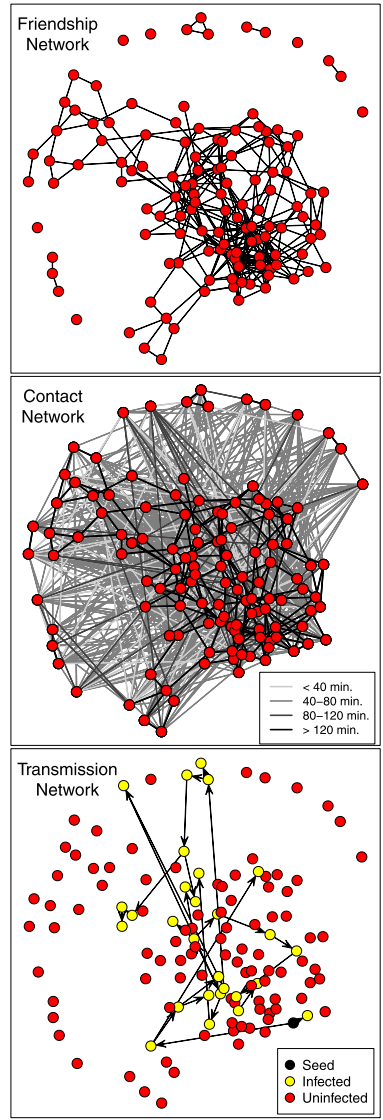}

\caption{The top figure shows a subset of the Add Health friendship
network. The middle figure shows a simulated contact network based on
this friendship network; here an edge represents one or more contacts
during one day, and the shade of gray represents the total duration of
contact between each pair. The bottom figure shows a simulated
transmission network based on this contact network. The seed of the
epidemic is black; the color of other nodes indicates whether they
became infected during the epidemic or not. The friendship network was
plotted with a standard layout algorithm which places connected
vertices closer and disconnected vertices farther away in order to
reduce numbers of edge crossings and reflect inherent symmetry
[Fruchterman and Reingold (\protect\citeyear{fruchterman})]. The other two
plots use the same vertex layout as the friendship network.}
\label{fignets}
\end{figure}

Figure \ref{fignets} shows the relationship between friendship network,
contact network and transmission network. The top panel depicts a
subset of the Add Health friendship network.
%f2 ###
%
The middle panel shows a simulated contact network among this same set
of students for one day. Here, an edge between two nodes means they
made one or more contacts, and the shade of the edge represents the
total duration of contact throughout the day for that pair. The contact
network is denser than the friendship network, as students tend to
contact their close friends as well as many other students during a~typical school day.
Of key scientific interest is the transmission
network, an example of which is shown in the bottom panel. The
dependency in the networks is shown by the higher numbers of contacts
between friends and higher numbers of transmission events between
friends. In this paper we focus on inference of the contact network and
explore how contact network structure impacts the transmission process.

%s3 ###
\section{Methodology}
\label{sectionmethods}

Our friendship network data forms the basis for our contact network
model. One approach to model the contact network for these students
would be to let friendships represent contacts, assuming that students
contact all of their close friends, and no other schoolmates, on
a~given day. Such a model would be overly simplistic. We believe that
students are more likely to contact their friends and make longer
durations of contacts to friends, but also contact other students in
their classes and in the school. We build a complex model capturing
these tendencies. We model contact behavior among the students in the
Add Health friendship network, using the epidemic survey to estimate
numbers of contacts and preference of contacts to friends. Finally, we
estimate the friendship network itself from individual-level attributes
so that our model can be used for an arbitrary school.

%s3.1 ###
\subsection{Modeling the contact network conditional on the friendship network}
\label{subsectionoutline}

We first describe our methodology to estimate the contact network
conditional on the empirical friendship network. We chose to model the
friendship network itself as a final step. Comparison of epidemics
based on the empirical friendship network to those based on a
friendship network simulated from our model assists with model
validation for the friendship network model. Through this comparison,
we assess whether the friendship network model captures the network
structures relevant to the transmission process.

We can represent the contact network graphically by letting each
student be a node and each contact be an edge between two nodes. The
degree of a node is the number of contacts made by that student during
one day. We denote the contact network by an $n$ by $n$ sociomatrix
$Y$, where $n$ is the number of students in the school. $Y_{ij}$
denotes the number of 10-minute contacts between student $i$ and
student $j$. Each pair of nodes in the network is referred to as a dyad.

We assume that students have seven classes of 40 minutes each, a
50-mi\-nute lunch break and five 10-minute nonlunch breaks. We define a
``contact'' to be a 10-minute face-to-face social contact. If two
students spend an hour together, that is considered six ``contacts.''
We allow a maximum of 38 contacts (6 hours and twenty minutes) between
any two students on a given~day.

%s3.2 ###
\subsection{Modeling the contact degree distribution}
\label{subsectiondegrees}

We model the degree distribution of the network using data from the
epidemic survey. We assume that students reported numbers of
schoolmates contacted rather than numbers of 10-minute chunks of time
spent in contact. We model break contacts and lunch contacts with
negative binomial distributions because the observed sample mean and
variance %(6 and 54 for break contacts; 11 and 154 for lunch contacts)
indicate over-dispersion. We used number of friends as a predictor,
expecting students with higher numbers of friends to make more contacts
at school. We fit a generalized linear model with the glm.nb function
in the MASS library in R [\citet{mass}, \citet{R}]. Before fitting, we modified
some outliers: we recoded 11 reports of break contacts greater than 20
to 20, and we removed 11 reports of numbers of close friends that were
over 40, assuming that these students defined ``close friend''
differently than the others. Our model estimates a mean of 4.5 break
contacts for a student with zero friends and an increase in expected
number of break contacts by a factor of 1.03 for each additional friend
(95\% C.I.: [1.01, 1.04]). Using the same model, we found no
association between lunch contacts and number of close friends; the
model estimated an increase in expected number of lunch contacts by a
factor of 1.00 for each additional friend (95\% C.I.: [1.00, 1.00];
$p=0.32$). Therefore, we estimated the lunch contact distribution with a
negative binomial distribution with no predictor. To reduce the
influence of outliers, we used a fitting procedure which assumes that
reports above a specified cutoff contain no other information apart
from being above the cutoff. We chose a cutoff of 30 for average lunch
contacts, so reported lunch contacts over 30 were treated as if these
students had reported ``$>$30'' lunch contacts. We assumed that
lunch
contacts could be 10, 20, 30, 40 or 50 minutes with equal probability,
so we multiplied each simulated contact by a randomly chosen number
between one and five. The fitting procedure is implemented with the
\texttt{anbmle()} function available in the \texttt{degreenet} package in
R [\citet{R}, \citet{degreenet}].

Classroom contacts were not reported, so we create a model for the
within-classroom contact degree distribution as follows. We assumed
students take classes only with others in the same grade. Each student
is randomly assigned to have 2, 3 or 4 class neighbors with
probabilities $1/9$, $4/9$ and $4/9$ in each class. We assumed that students
make 40 minutes of continuous contact with each of their neighbors
during each class period, that they have the same class neighbors each
day, and that they only contact class neighbors during class time.

The distribution of total contacts is obtained by summing the classroom,
lunch and break contacts together for each student. This distribution
has a~mean of 148, or 25 person-hours of contact per student per day. We
validated our fitted degree distribution by comparing it to contact
reports from an alternate data source, the POLYMOD study [\citet
{mossong}]. This validation is described in the supplementary
material [\citet{suppA}].

%s3.3 ###
\subsection{Modeling the contact network conditional on the degrees}
\label{subsectionlink}

We depict the degrees as a set of nodes representing students, each of
whom has a~number of stubs representing their contacts. In this section
we describe how these stubs will be linked, forming the entire network
of contacts between students. We denote the degrees as a vector $D$ of
length $n$, where $D_i$ is the number of contacts student $i$ makes in
one day.

Let $Y_{\bl}$ be the sociomatrix of contacts occurring during any of the
class breaks or during lunch and $Y_c$ denote the within-class contact
sociomatrix, so $Y=Y_{\bl}+Y_c$. We model $Y_{\bl}$ conditional on the
break and lunch contact degrees, and we model $Y_c$ conditional on the
class contact degrees. Let $D_{\bl}$ denote the vector of break/lunch
degrees. Then the probability distribution for $Y_{\bl}$ can be expressed:
\[
\mathrm{P}(Y_{\bl}=y_{\bl}) = \sum_{d_{\bl}}\mathrm
{P}(Y_{\bl}=y_{\bl}|D_{\bl}=d_{\bl})\mathrm{P}(D_{\bl}=d_{\bl}).
\]

Because respondents in the epidemic survey report an average of 68\% of
contacts occurring to friends, our model distributes 68\% of contacts
to friends and 32\% to nonfriends, with a maximum of 10 contacts per
dyad allowed (since there are 100 minutes in the 5 breaks plus lunch
period combined). Apart from these constraints, contacts occur randomly
conditional on the degree distribution,\vadjust{\goodbreak} which means that all networks
satisfying these constraints have equal probability:
\[
P(Y_{\bl}=y| D_{\bl}=d_{\bl})=
\cases{
\displaystyle \frac{1}{c(d_{\bl})}, &\quad if $\displaystyle \frac{ \sum_{\{i,j\dvtx i,j\
\mathrm{are}\ \mathrm{friends}\}}y_{ij} } {\sum_{i,j} y_{ij}}=0.68$\vspace*{2pt}\cr
&\quad and $y_{ij}\le10$ \vspace*{2pt}\cr
&\quad and $\displaystyle \sum_j y_{ij} = d_{\bl,i}$\vspace*{2pt}\cr
0, &\quad otherwise,}
\]
where $c(d_{\bl})$ is a normalizing constant.

We develop a method to simulate networks from a specified degree
vector, with random mixing conditional on degree and permitting
multiple edges (up to a specified maximum) between two nodes. Our
method is an extension of the \texttt{reedmolloy()} function in the
\texttt{degreenet} package in R [\citet{R},
\citet{degreenet}]. Denote the maximum number of edges~$m$ and the
target percentage of edges to friends $p$, and let $d_i$ denote the
degree of node $i$. We first compute the target number of contacts
between friends, denoted by $T$:
\[
T=p \frac{\sum_{i=1}^n d_i}{2}.
\]
We randomly sample a stub, and let $i$ denote the node possessing this
stub. We consider the set of friends of $i$ which have fewer than $m$
edges to $i$. We randomly sample one friend from this set, with
probability proportional to the remaining (unassigned) degree of each
friend. Then the two stubs are connected. This procedure is repeated
until we have $T$ contacts between friends. Next, we repeat the process
for nonfriend contacts. The procedure requires the sum of the degrees
to be even and enough friendships so that $m$ times the number of
friendships is greater than or equal to $T$. Since self-self edges are
not permitted, the procedure also requires
$\max(\mathbf{d})\le\sum_{\{i\dvtx d_i\le
\max(\mathbf{d})\} } \min\{m, d_i\}$.

To simulate break/lunch contact networks, we first sample lunch and
break contacts from the fitted degree distributions for each student.
Then we distribute 68\% of contacts to friends, with a maximum of 10
contacts occurring between any pair of friends.

Next we describe the probability distribution for our class contact
network. We assume that students take classes only with others in the
same grade. We model the matrix of class neighbors, $Y_{\mathrm
{neighbors}}$, where $Y_{\mathrm{neighbors}, ij}$ is the number of
classes in which $i$ and $j$ are neighbors. We then assume that each
pair of class neighbors makes 40 minutes of continuous contact during
each shared class, so the contact matrix is $Y_c = 4 Y_{\mathrm{neighbors}}$.

To model $Y_{\mathrm{neighbors}}$, let $Y_k$ denote the $n$ by $n$
matrix showing classroom neighbors within grade $k$. That is, if $i$
and $j$ are in grade $k$, then the $ij$th element of $Y_k$ is the
number of classes in which $i$ and $j$ are class neighbors, and if $i$
or $j$ is not in grade $k$, then $Y_{k,ij}=0$. Then $Y_{\mathrm
{neighbors}} = Y_7 + Y_8 + \cdots+ Y_{12}$. We model degrees of class
neighbors within grade $k$ as described previously. Because 74\% of
respondents in the epidemic survey reported sitting next to ``A mix of
friends and nonfriends'' in class, we assume that 50\% of class
neighbors are friends. Using the procedure described above, we
distribute 50\% of class neighbors to be friends and allow students to
be neighbors in more than one class, with a maximum of 7 shared
classes. Thus,
\[
P(Y_k=y| D=d)=
\cases{
\displaystyle \frac{1}{c(d)}, &\quad if $\displaystyle
\frac{\sum_{\{i,j\dvtx i,j\ \mathrm{are}
\ \mathrm{friends}\}}y_{ij} }
{\sum_{i,j} y_{ij}}
=0.50, y_{ij} \le7,$ \vspace*{2pt}\cr
&\quad and $\displaystyle \sum_j y_{ij} = d_i$,\vspace*{2pt}\cr
0, &\quad otherwise,}
\]
where $c(d)$ is a normalizing constant.

To simulate a class contact network for one day, we first sample class
neighbor degrees for each grade from the fitted degree distributions.
Then we use our modified reedmolloy() function to distribute 50\% of
neighbors to friends, allowing two students to be neighbors in a
maximum of 7 classes, for each of the grades. We multiply these class
neighbor matrices by four to obtain class contact matrices for each
grade, and sum the seven grade-specific class contact matrices to
obtain the class contact matrix for the entire day.

%s3.4 ###
\subsection{Contact network simulation procedure}
\label{subsectioncontactsimulation}

In this section we describe our algorithm to simulate contact networks
from our model. The uncertainty in estimation of the input parameters
to our model will propagate to create uncertainty in epidemic
predictions. We use a nonparametric bootstrap to estimate this
uncertainty [\citet{efron}].

We simulate a contact network as follows:

\begin{longlist}[(2)]
\item[(1)] Resample with replacement from the epidemic survey.
\item[(2)] Using the resampled data, estimate degree distribution parameters
(as described in Section \ref{subsectiondegrees}), and compute the
average percentage of contacts to friends. Denote this percentage by $X$,
where $E[X] = 68$\%.
\item[(3)] Simulate break and lunch contact degrees from the fitted distributions.
\item[(4)] Link stubs (as described in Section \ref{subsectionlink}) so that
$X$\% of break and lunch contacts are between friends.
\item[(5)] Simulate class neighbor degrees from the assumed degree
distribution, described in Section \ref{subsectiondegrees}.
\item[(6)] Link stubs (as described in Section \ref{subsectionlink}) so
that 50\% of class neighbors are friends.
\item[(7)] Multiply by 4, assuming that class neighbors make 40 minutes
of continuous contact in each shared class.
\item[(8)] Sum the break/lunch contact network and class contact
network matrices to obtain the contact network matrix for one day.
\end{longlist}

To produce a dynamic contact network model, we sample a new break/\allowbreak lunch
contact network each day of the
influenza season, but keep the same class contact network throughout
the influenza season. In the supplementary material, we present
descriptive analyses of contact networks simulated from our model and
find their properties to be consistent with our observed data
[\citet
{suppA}]. % We also provide a flow chart of our algorithm.

%s3.5 ###
\subsection{Influenza simulation procedure}
\label{subsectionsimulation}

We simulated influenza outbreaks in schools using the natural history of
influenza as was done by \citet{dennis}. We assume that each
student has
an incubation period (time between exposure and appearance of symptoms)
of 1, 2 or 3 days with probabilities 0.30, 0.50 and 0.20, respectively.
Each infected person stays infected for exactly
6~days, after which he/she is moved to the immune category.
Transmission can
occur only when contact is made between an infected person and a susceptible
person. For each infected person, we sample a curve of viral load over time
from those of six patients in the human challenge study described
in \citet{murphy} and \citet{baccam},
and we assume that the infectiousness of each person on a given
day is proportional to their viral load. We assume that 67{\%} of students
become symptomatic during their infectious period, and symptomatic
people are twice as infectious as asymptomatic people, so their
infectiousness is proportional to twice their viral load. Let $p_{t,i}$
denote the per-10-minute transmission probability of person $i$ on day
$t$. The events that $i$ transmits to~$j$ during two different
10-minute contacts are dependent, since transmission during the earlier
contact precludes transmission during the latter. Thus, if~$j$ is susceptible,
\[
\mbox{P($j$ escapes infection by person $i$ on day $t$)} =
(1-p_{t,i})^{Y_{ij}},
\]
so
\[
\mbox{P($j$ infected on day $t$)} = 1 - \prod_{i=1}^n
(1-p_{t,i})^{Y_{ij}}.
\]

We assume 75{\%} of sick students withdraw to the home: 20.3{\%} on the
first day they have symptoms, 39.7{\%} on the second, and 15{\%} on
the third [\citet{dennis}, \citet{elveback}].

We used mean per-10-minute transmission probabilities ranging from
0.001 to 0.007. We track the epidemic until no infected people remain.
We estimated the probability of epidemic (defined as more than 200
students becoming infected), the peak date of the disease season and
the final epidemic size. In\vadjust{\goodbreak} performing simulations for model comparison
(described in the following section), we simulated 500 outbreaks for
each model; this number was sufficient to distinguish between them. In
performing simulations validating the friendship network model fit, we
simulated 2,000 outbreaks, which was sufficient to validate model fit.
For simulations using our final model and random mixing, with and
without interventions, we simulated 10,000 outbreaks for each scenario
to minimize uncertainty in epidemic outcome estimates.

%s3.6 ###
\subsection{Model comparison}
\label{subsectionselection}

We compared three different versions of the contact network model. In
the \textit{dynamic contact network model}, students keep the same
class contacts for the duration of the influenza season, but we sample
a new break/lunch contact network each day. There is, to our knowledge,
no previous work on modeling dynamic within-school contact networks,
and we consider this to be our most realistic model. To assess whether
these dynamics influenced epidemic predictions, we compared this to a
\textit{static contact network model}, in which students contact the
same people each day for the duration of the influenza season. The
static network approach is commonly used to model influenza
epidemics [\citet{miller}]. Finally, we investigated whether the
transmission process is driven purely by the friendship structure by
implementing a \textit{friendship-only} model, in which students only
contact their friends. We calibrated the friendship-only model so that
the expected total number of contacts in all models is the same.
Comparison to this model will reveal whether the additional network
structure we added, including a proportion of contacts to nonfriends,
variation in contact degree and classroom structure, has an impact on
epidemic predictions.

\begin{figure} % float placement: (h)ere, page (t)op, page (b)ottom,
%other (p)age

\includegraphics{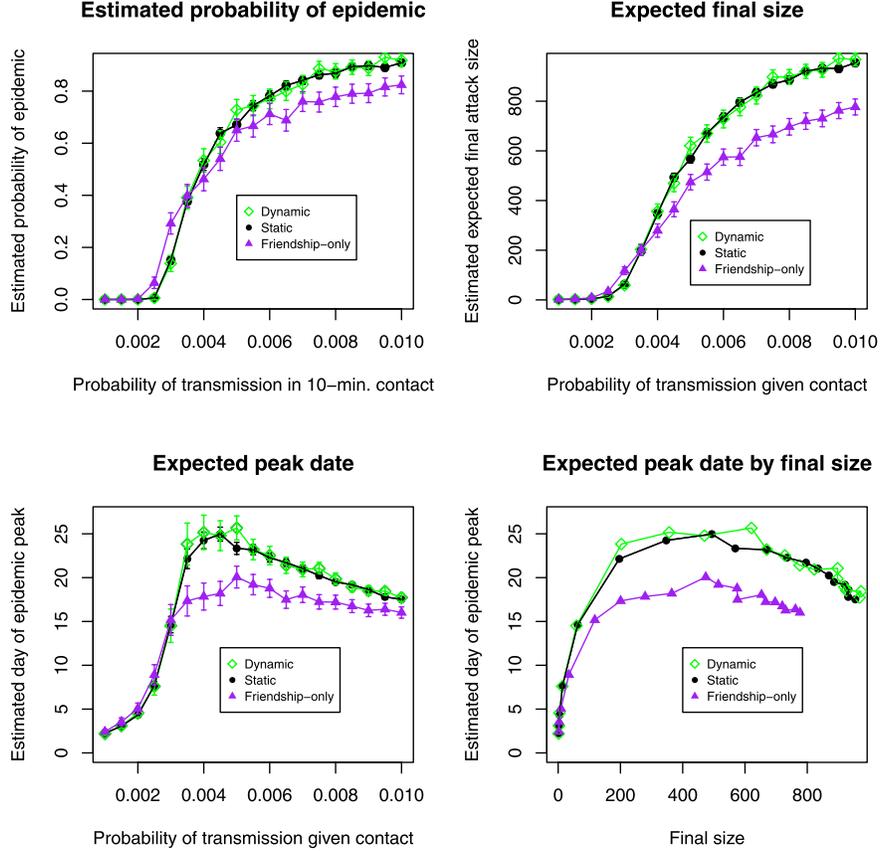}

\caption{Comparison of epidemic outcomes for three different contact
network models, based on 500 simulated epidemics for each contact
network model.}
\label{figcomparemodels}
\end{figure}

We simulated 500 epidemics over each of these three models using the
natural history of influenza described above. Epidemic outcomes,
displayed in Figure~\ref{figcomparemodels}, are essentially identical
in the static and dynamic contact network models.
A similar result was found in a different setting
by \citet{Sthetal11}.
%f3 ###
%
This is because our dynamic model creates a sequence of highly
correlated contact networks. Although break/lunch contact networks are
sampled independently from one day to the next, these networks are
dependent because they rely on the same underlying friendship network,
which stays the same for the whole influenza season. We found that most
contacts which change status from on to off or vice versa are only 10
minutes in duration. These dynamics do little, if anything, to shift
the course of the epidemic. The friendship-only model behaves quite
differently. The friendship-only model is oversimplified, and the
additional network structure of classroom contacts and distribution of
nonfriend contacts creates a~more realistic model. Therefore, we
selected the static network model for our final model.

%s3.7 ###
\subsection{Modeling the friendship network}
\label{subsectionergm}

Our contact network model described above is conditional on the
empirical friendship network. To generalize our model, we need to
model\vadjust{\goodbreak}
the friendship network itself; we do so using an exponential family
random graph model (ERGM). We represent the friendship network by a
sociomatrix $Y$. An ERGM models the sociomatrix for a network of fixed
size as follows:
\[
P(Y=y|\mathbf{\theta}) = \frac{
e^{\sum_{i=1}^k \mathbf{\theta}^T \mathbf{g(y)} }}{\kappa(\mathcal{Y})}.
\]
Here, $\mathcal{Y}$ denotes the space of all possible networks of this
size, and $\kappa(\mathcal{Y})$ is a~normalizing constant which
ensures that the probability distribution sums to 1. $\mathbf{\theta}$
is a vector of parameters, and $\mathbf{g(y)}$ is a vector of network
statistics, such
as the number of edges between actors of the same race, the number of
triangles or others. These statistics capture social principles such as
the tendency to befriend others with like attributes or transitivity. A
dyad-independent ERGM is a model in which the probability of observing
an edge on one dyad is independent of the probability of observing an
edge on other dyads (although it may depend on individual-level and
dyadic attributes). The parameter estimates are obtained by their
maximum likelihood estimates (MLE). In many cases there is no analytic
form for the normalizing constant $\kappa(\mathcal{Y})$, which is
difficult to approximate because of the large number of possible
networks for an undirected network. Instead the MLE is approximated
through a Markov chain Monte Carlo procedure described by \citet{geyer}.
However, a dyad-independent ERGM may be estimated with logistic
regression rather than the MCMC procedure.

%s3.7.1 ###
\subsubsection{Model selection}

Our model is based on the work of \citet{goodreau}, who use exponential
random graph models to describe friendship patterns in all 80 schools
in the Add Health data set. The authors model the network of mutual
friendship nominations for each school. Their model includes sociality
terms for each grade, race and gender, selective mixing by race, grade
and gender, and a transitivity term which captures the tendency of
friends of friends to also be friends, conditional on other terms in
the model. Our ERGM includes these effects minus the transitivity term,
so is slightly simpler, although we also included a school mixing effect.

Table \ref{tabcoefests} shows coefficient estimates for our model. The
sociality terms capture whether 8th graders form larger numbers of
friendships, on average, than seventh graders (the reference category
for grade), etc. These terms are interpreted as follows: a friendship
is $\exp(0.54)=1.71$ times more likely to occur from a randomly chosen
person to an eighth grader than to a seventh grader, assuming that the
eighth grader and seventh grader are identical on other attributes
included in the model. Other sociality terms are interpreted similarly.
We see, for example, that eighth graders are significantly more social
than seventh graders, but twelfth graders are not. Mixing coefficients
represent the tendency to form friendships with others who have the
same attributes as oneself; these are interpreted as follows: a
friendship between two seventh graders is $\exp(2.3)=9.9$ times more
likely to occur than a~friendship between two students in different
grades, all other attributes being equal. The coefficient is $-\infty$
for the race missing category because there are no friendships among
this very small ($n=11$) group of students.

%t1 ###
%
\begin{table}
\caption{Coefficient estimates for Exponential Family Random Graph
Model (ERGM) fitted to the Add Health friendship network. Significance
levels are denoted as follows:\break ***($p\le0.001$), **($p\le0.01$), *($p\le
0.05$) and ****($p\le0.1$)}
\label{tabcoefests}
\begin{tabular*}{\tablewidth}{@{\extracolsep{\fill}}lrccc@{}}
\hline
&&&&\multicolumn{1}{c@{}}{\textbf{Significance}}\\
\multicolumn{1}{@{}l}{\textbf{Variable}} & \multicolumn{1}{c}{\textbf{Coef. (SE)}}
& \multicolumn{1}{c}{\textbf{Significance}}
& \multicolumn{1}{c}{\textbf{Factor}}
& \multicolumn{1}{c@{}}{\textbf{of factor}} \\
\hline
Edges sociality& $-$10.91 (0.78)&***\\
\quad Grade 8 & 0.54 (0.13)&*** &Grade &***\\
\quad Grade 9 & 0.24 (0.09)&**\hphantom{*}\\
\quad Grade 10 & 0.57 (0.09)&***\\
\quad Grade 11 & 0.45 (0.09)&***\\
\quad Grade 12 & $-$0.01 (0.09) &\\
\quad Black & 0.12 (0.10) & &Race &***\\
\quad Hispanic & 0.81 (0.09)&***\\
\quad Asian & $-$0.19 (0.12)&.\hphantom{**}\hspace*{2pt} \\
\quad Mixed race & 0.71 (0.09)&***\\
\quad Race missing & 0.58 (0.14)&***\\
\quad Male & 0.3 (0.03)\hphantom{0}&*** &Sex &***\\[4pt]
Selective mixing & &\\
\quad School & 1.73 (0.07)&*** &School &***\\
\quad Male & 1.05 (0.38)&**\hphantom{*} &Sex &**\hphantom{*}\\
\quad Female & 1.18 (0.38)&**\hphantom{*}\\
\quad Grade 7 & 2.3 (0.15)\hphantom{0}&*** &Grade &***\\
\quad Grade 8 & 1.51 (0.15)&***\\
\quad Grade 9 & 1.88 (0.11)&***\\
\quad Grade 10 & 1.17 (0.11)&***\\
\quad Grade 11 & 1.61 (0.12)&***\\
\quad Grade 12 & 2.71 (0.13)&***\\
\quad White & 1.03 (0.10)&*** &Race &***\\
\quad Black & 3.19 (0.16)&***\\
\quad Hispanic & $-$0.5 (0.33)\hphantom{0}& \\
\quad Asian & 2.94 (0.26)&***\\
\quad Mixed race & $-$0.58 (0.20)&**\hphantom{*}\\
\quad Race missing & \multicolumn{1}{c}{$-$Inf (NA)} &\\
\hline
\end{tabular*}
\end{table}

We assessed whether our model captures the relevant network structures
by simulating friendship networks from our estimated model parameters,
simulating contact networks based on the simulated friendship data (as
described in Sections \ref{subsectionoutline}--\ref{subsectionlink}),
and then simulating 2,000 influenza epidemics over these contact
networks (as described in Section \ref{subsectionsimulation}). If our
friendship model is adequate, epidemic outcomes from these simulations
should resemble those estimated in simulations based on the empirical
friendship network. We performed this procedure for three different
simulated networks from our ERGM.

%s3.8 ###
\subsection{Methodology to compare contact network model to random mixing}
\label{subsectioncompare}

We simulated influenza epidemics over the static contact network model and
compared them to simulations over a random mixing scenario. We
calibrated the
random mixing model so that the expected number of people contacted per
student per day is the same as in the friendship-based model (36), and the
duration of contact is equal to the average duration of contacts in the
friendship-based model (41~minutes).

We first simulated epidemics with no intervention. Then we simulated a
\textit{reactive grade closure} intervention, in which the entire grade
of a student
manifesting influenza symptoms is closed one day after detection of
symptoms. Next, we investigated the impact of network structure on the
estimated effect of a \textit{targeted antiviral prophylaxis} (TAP)
strategy. Under
this strategy, all symptomatic people are given five days of antiviral
treatment, and their contacts are given ten days of antiviral prophylaxis,
starting the day after symptoms appear. Based on estimates by
\citet
{halloran2007},
we assume an antiviral efficacy against susceptibility ($\mathit{AVE}_{S}$)
of 63{\%}, antiviral efficacy against infectiousness ($\mathit{AVE}_I$)
of 15{\%}, and antiviral efficacy against pathogenicity ($\mathit{AVE}_P$)
of 56{\%}. Thus, the probability of getting infected during
one contact is reduced by a factor of $1-\mathit{AVE}_S = 0.37$ if the susceptible
person is receiving prophylaxis, and further reduced by a factor of
$1-\mathit{AVE}_I=0.85$
if the infectious person is receiving antiviral treatment.
Treated people are $1-\mathit{AVE}_P=0.44$ times less likely to become
symptomatic than untreated people.

%s4 ###
\section{Results}
\label{sectionresults}

\begin{figure} % float placement: (h)ere, page (t)op, page (b)ottom,
%other (p)age

\includegraphics{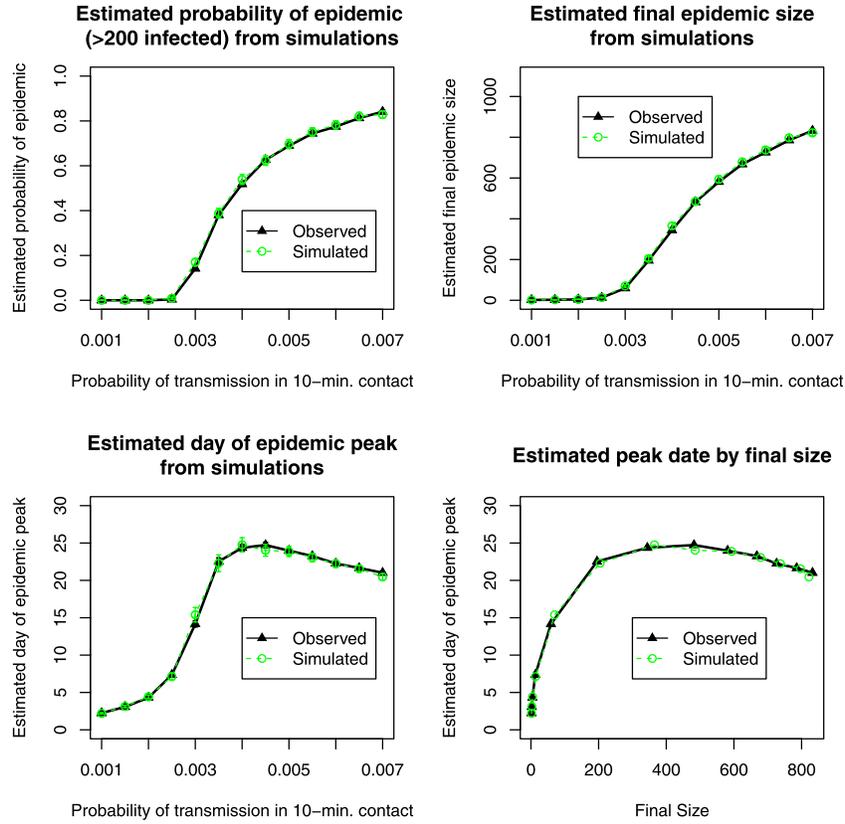}

\caption{Comparison of epidemic outcomes from simulations based on the
observed friendship network to those based on a friendship network
simulated from our friendship network model.}
\label{figdyaddep}
\end{figure}

Figure \ref{figdyaddep} compares epidemic outcomes for simulations
based on the empirical friendship network to those based on the
simulated friendship network. The results are nearly identical,
indicating that our estimated friendship network model captures the
network structures relevant for disease transmission. We display
epidemic outcomes for transmission probabilities in range displaying a
broad spectrum of epidemic possibilities: 0.001 to 0.007. Transmission
probabilities smaller than 0.002 were too small to produce epidemics,
so the probability of epidemic is zero for that range, while estimated
final size and peak date are negligible compared to estimates for
larger transmission probabilities. The error bars in all plots in this
section depict uncertainty arising both from estimation of parameter
inputs to our model, as well as from the simulations. In most cases,
the width of the error bar is smaller than the plotting symbol.

%f4 ###
%

%
%f5 ###
%
\begin{figure}

\includegraphics{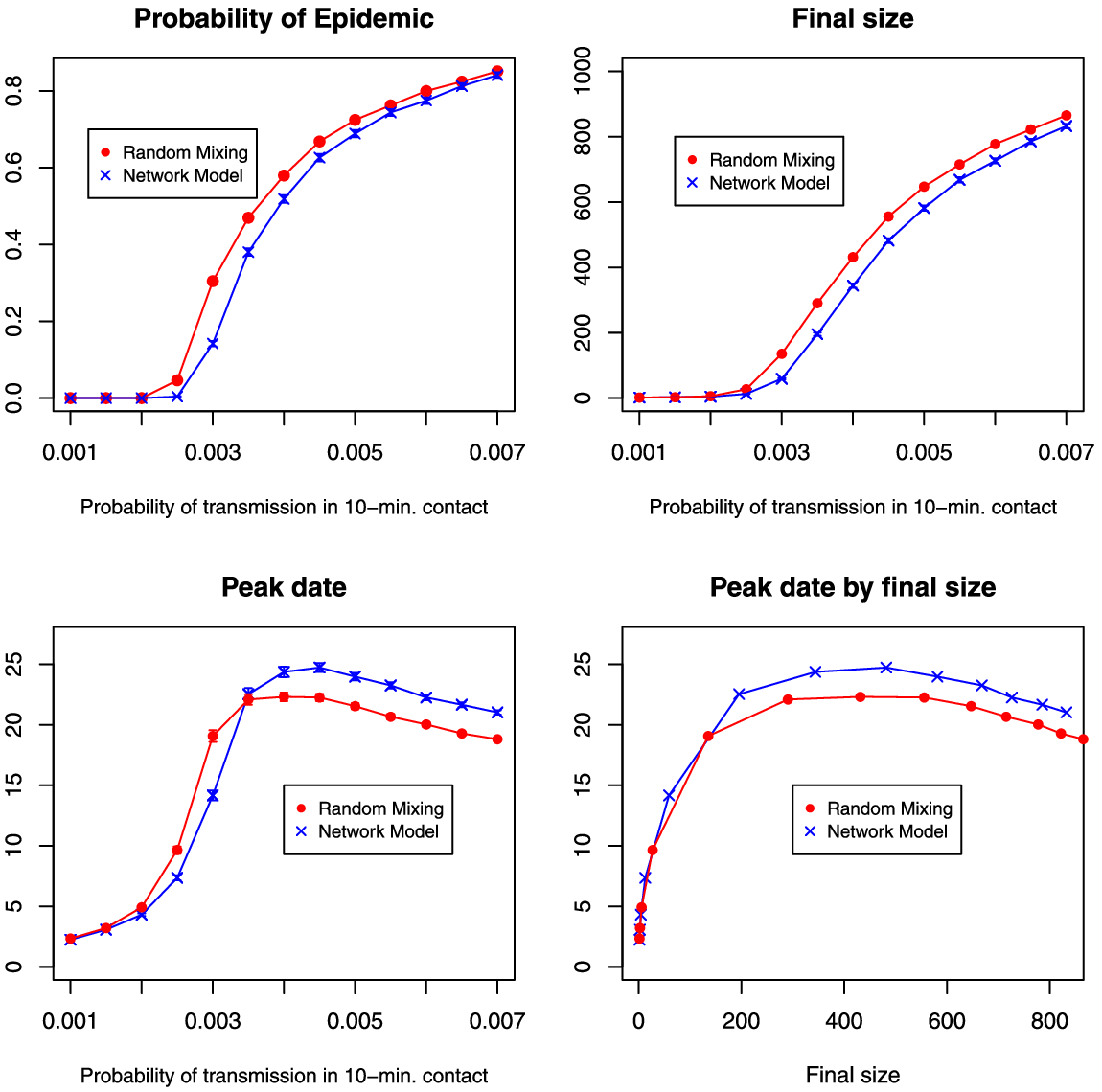}

\caption{Comparison of epidemic outcomes from simulations over the
static contact network model to those assuming random mixing.}
\label{fignointervention}
\end{figure}

Figure \ref{fignointervention} compares epidemic outcomes for
simulations over the static contact
network model to those from simulations performed over a random mixing
scenario. The estimated probability of epidemic and final size are smaller
in the contact network model than in a random mixing model. The repetition
in contacts in our network model reduces the pool of susceptibles accessible
to an infected person, who continues to contact people he/she has already
infected. The transitivity present in friendship patterns further
limits the
potential for disease spread. Friends are likely to have mutual
friends, so
the set of susceptible friends of an infected person is reduced by
transmission from other mutual friends. Figure \ref{fignointervention}
also shows the estimated
peak date of the disease season: the day with the largest number of infected
students. For probabilities of transmission under 0.0035, the epidemic peaks
sooner under the network model; for higher probabilities of transmission,
the epidemic peaks later. The threshold value occurs because the
relationship between peak date and transmission
probability is confounded by final size. The plot of peak date by final size
shows that the network model peaks later for all final sizes than a random
mixing model. The spread of the virus is slowed by the clustering and
repetition in contacts in the network model.

%f6 ###
%
\begin{figure}

\includegraphics{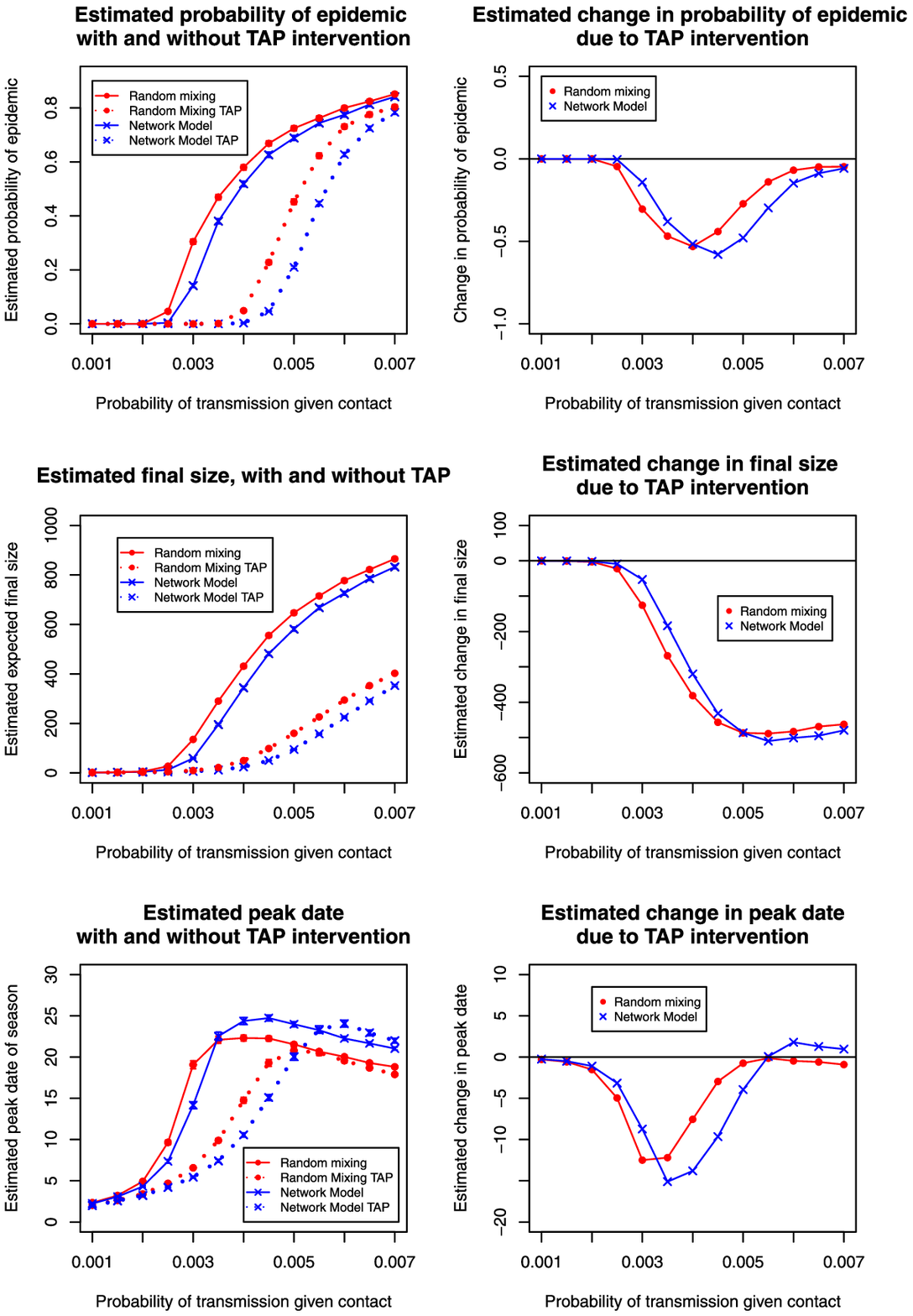}

\caption{Estimated effect of targeted antiviral prophylaxis (TAP)
intervention on probability of epidemic, final size and epidemic peak
date under the static contact network model compared to random mixing.}
\label{figTAPall}
\end{figure}

The top row of Figure \ref{figTAPall}
shows the estimated probability of an epidemic with targeted
antiviral prophylaxis intervention under the network model and the random
mixing model and the change in estimated probability of epidemic under
both scenarios.%%
These plots describe the estimated effectiveness of this intervention
for containing the epidemic.
Under both scenarios, the probability of epidemic is reduced to zero for
transmission probabilities under 0.0035. If we were using either model
for prediction, the right-hand plot would be the relevant one, and for
this range of transmission probabilities, random mixing estimates a
larger improvement than the network model. For example, when the
transmission probability is 0.003, random mixing estimates a reduction
of 0.30 in probability of epidemic, while the network model estimates
this reduction to be 0.13. At transmission probabilities above 0.0035,
the estimated probability of epidemic is higher under the random
mixing model than the network model. This strategy is more effective under
the network model because the people prioritized for prophylaxis are those
who are repeatedly exposed through daily contact to infectious
individuals. In the random mixing model, the contacts of an infectious
person on one day are unrelated to his contacts on the following
day, so the prioritization of antiviral to contacts has no effect.

The second row of Figure \ref{figTAPall} shows a similar pattern with
final size, but with a threshold value of 0.005 instead of 0.004.
The third row shows substantial differences in peak date predictions
between the two models. A~delay in peak date helps the public health
department develop a response to the epidemic. However, the epidemic
peaks earlier with the intervention under both scenarios. This is
because the relationship between peak date and transmission probability
is confounded by final size; both interventions reduce the final size
drastically, so the (much smaller) peak occurs sooner.

In simulating the TAP intervention, we distributed antiviral
prophylaxis to all (100\%) contacts of symptomatic students, thus
assuming that symptomatic students would accurately recall and report
100\% of the students they contacted on their first day of symptom
onset. In reality, students may recall only a subset of the people they
contacted on their first symptomatic day. To assess the impact of this
assumption, we repeated the analyses assuming
that students reported only 90\% of contacts, and again assuming that
they reported only 75\% of contacts. These results are included in the
supplementary material [\citet{suppB}]. These different scenarios only slightly
shifted the results, maintaining our qualitative and quantitative findings.

\begin{figure}

\includegraphics{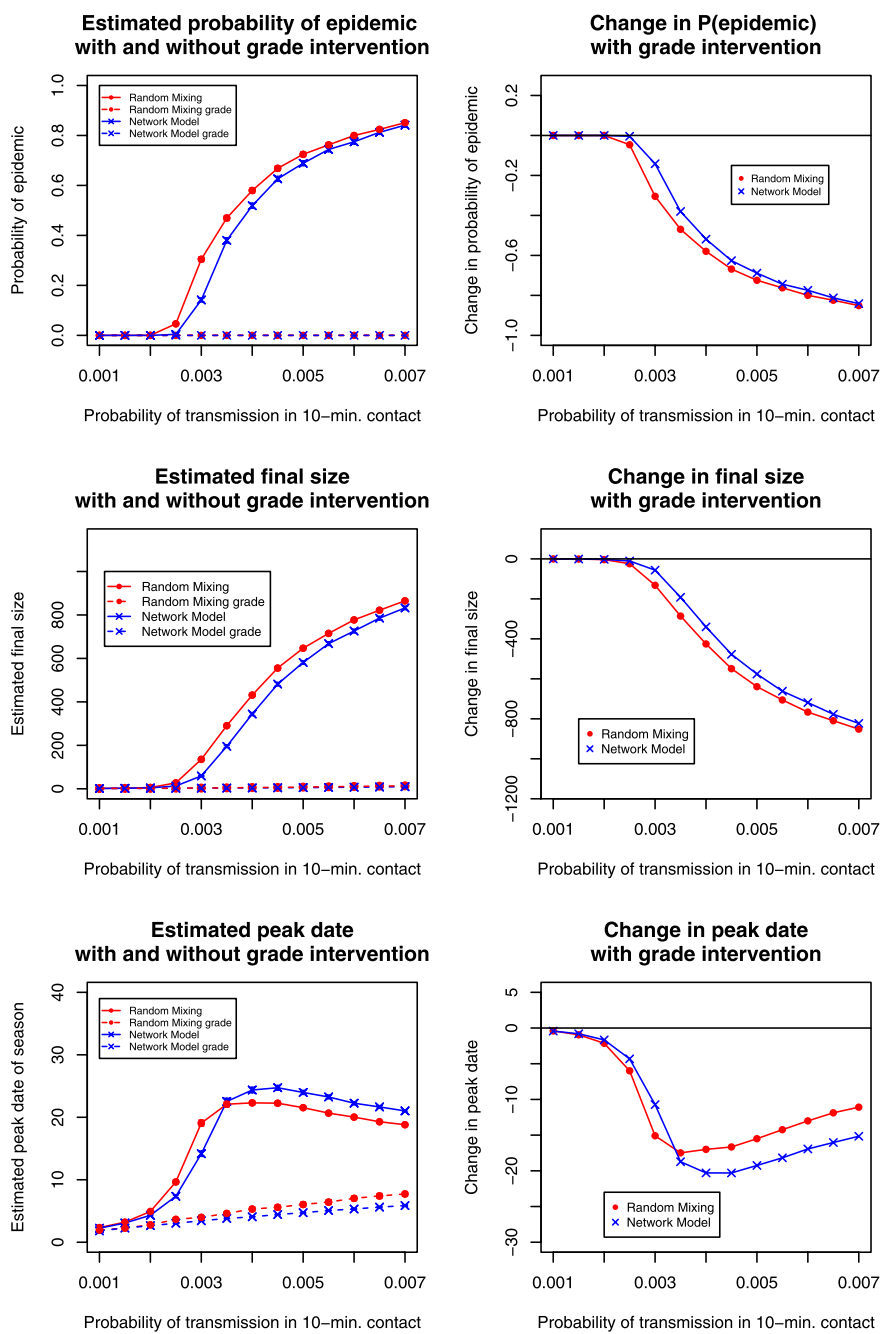}

\caption{Estimated effect of reactive grade closure intervention on
probability of epidemic, final size and epidemic peak date under the
static contact network model compared to random mixing.}
\label{figgradeall}
\end{figure}

The first row of Figure \ref{figgradeall} shows that under both models
the grade closure strategy reduces the probability of epidemic to zero
for all transmission probabilities. %%
Since grade closure is expensive on a societal level, our
model could be used to perform cost-effectiveness strategies, where the cost
of grade closure is weighed against the severity of the influenza
strain and
its societal impact. The right-hand plot in the second row of
Figure \ref{figgradeall} shows that if we were
willing to use grade closure once the reduction in probability of
%f7 ###
%
epidemic exceeded a threshold value (e.g., 0.20), the cutoff
transmission probability
would be different under the two models. The third row shows
differences in peak date predictions under the grade closure strategy.

%s5 ###
\section{Discussion}
\label{sectiondiscussion}

Our work in this paper yields three broad findings. First, our realistic,
data-driven contact network model produces substantially different estimates
of epidemic outcomes and intervention effectiveness than a~random
mixing scenario, and the differences vary by transmission probability.
Second, we found evidence that in a high school setting, a static
contact model is sufficient to characterize epidemic progress. However,
our dynamics in contact behavior occurred only during class breaks, so
relied on the assumptions that within-classroom seating configurations
are constant over time and that interaction occurs only with one's
immediate class neighbors within each class. We recommend collecting
dynamic contact data and further investigating the hypothesis that
dynamic networks and static networks produce similar epidemic
predictions. Once dynamic, within-class contact reports are obtained,
we can integrate this information into our model and test our
hypothesis that a static contact network adequately represents the
contact behavior relevant for epidemic predictions. Third, a
dyad-independent ERGM adequately captures the friendship network
structure relevant to the disease transmission process. The
dyad-independent model is advantageous, as its parameters can be
estimated with logistic regression instead of relying on MCMC. Another
advantage of this model is that the probability of friendship depends
only on individual-level attributes,
so survey data on attributes of respondents and their friends is sufficient
to characterize the network.

Our model stands out from other epidemic simulation models for three
reasons. First, we infer the contact network using contact survey
reports, while others are not informed by contact surveys. Second, we
quantify uncertainty in predictions arising from uncertainty in
estimates of inputs to our model; this is not standard in the field.
Third, we validated our model by comparing the fitted degree
distribution to reports in an alternate data source and by comparing
joint and marginal distributions of variables of contact networks
simulated from our model to those in one of our data sources, the
epidemic survey.

Our work has several limitations. First, we have modeled contact and
transmission patterns in a single high school. The friendship patterns in
this high school may be different from those in other high schools,
especially schools of different sizes and racial compositions. We
hypothesize that in schools with different friendship structure, our key
findings that a dyad-independent ERGM is sufficient and that a static
contact model is adequate will still hold.

Another limitation of our work is that we have treated the Add Health
friendship data as complete rather than attempting to model the
unobserved friendship ties. Demographic information is unavailable for
nonrespondent students, and differences in demographics between
respondents and nonrespondents have not been studied. \citet
{markandkrista2} compared network characteristics of respondents to
nonrespondents in a different Add Health school, and found slight
differences, for example, that respondents received more friendship
nominations than nonrespondents. We found this pattern to hold in our
school as well: respondents received an average of~4.9 nominations
while the mean for nonrespondents was~3.5. However, if nonrespondents
are more likely to nominate other nonrespondents than respondents as
best friends, then the true means are closer together. Our work could
be extended by imputing demographics for nonrespondent students and
maximizing the likelihood obtained by summing over all possible values
for the missing edges [\citet{markandkrista}]. We consider our partially
observed friendship network to be a realistic representation of a~possible
friendship network and believe that correcting for missing
edges and attributes would only slightly impact our friendship network
estimates and would not substantively impact our epidemic outcome
estimates. Our main finding that a friendship-based contact model gives
rise to different estimates of epidemic outcomes than a random mixing
scenario is likely to hold with the complete friendship network.

Because Add Health respondents were limited to nomination of 5 friends
of each sex, there is truncation bias in the numbers of friends in the
friendship network. In this school, 86\% of respondents reported fewer
than 5 best male friends, 79\% reported fewer than 5 best female
friends and 95\% reported fewer than 10 best friends, so truncation
bias is relatively small. Students were instructed to list their
friends in order of closeness, so friendships that were truncated are
less close than the included ones. Moreover, by including nominators of
each respondent as friends even if they were not themselves nominated
by the respondent, we may have reduced the truncation bias. Because
this definition of friendship creates a degree distribution similar to
that collected in the epidemic survey, which had no truncation
mechanism (see Figure \ref{figdegreedist}), we expect any bias arising
from the
truncation in Add Health friendship reports to have minimal, if any,
impact on our results.

Reports in the epidemic survey are subject to a potentially high degree
of measurement error
because students were asked to estimate their average contact behavior. We
contrast this survey design to the POLYMOD study, in which respondents were
mailed paper diaries and instructed to carry them throughout a \mbox{24-hour}
period and record characteristics of each contact they made [\citet
{mossong}]. We
recommend a within-school POLYMOD type survey in which the students identify
their contacts from a school roster. We could directly model the contact
network from such a data set without inclusion of the friendship network
information. We believe that our model is the most realistic possible with
the available data, and the extent of measurement error is impossible to
determine without further studies. Proximity sensor data would also be less prone to measurement
error and can be used to characterize networks as in \citet{Sthetal}.

Another limitation of our model is that we did not incorporate data on
classroom contacts but rather created a model based on assumptions about
within-classroom contact behavior. A better understanding of classroom
contacts could be obtained by the POLYMOD-type within-school survey
described above, in which respondents include the time of day and whether
the contact occurs within a class. Further limitations include our
assumptions of perfect observation of symptoms and perfect reporting of
contact behavior during the targeted antiviral prophylaxis strategy,
but sensitivity analysis
demonstrated the latter assumption to have little effect.

We have modeled within-school contacts only. In reality, friends and
classmates also contact each other outside the school. We intend to expand
our school model to include all contacts between students in the school
occurring in all locations. The model we presented here is a natural first
step in building the expanded model.

We have developed a detailed, data-driven model of within-school social
contact behavior. We demonstrated that our network model predicts
different epidemic progress and intervention effectiveness than random
mixing, and we identified key network structures\vadjust{\goodbreak} influencing the
transmission process. We recommend further exploration into how network
structures influence the disease transmission process with the aim of
integrating network structure into epidemic models and simulators.

\section*{Acknowledgments}

We thank the Editor and Associate Editor of this paper for their
constructive comments and suggestions. We are very grateful to Martina
Morris for sharing the restricted-use Add Health data set, as well as
her feedback on this work. We are also grateful to Stephen Eubank,
Henning Mortveit and Madhav Marathe for sharing the epidemic survey
with us. We thank Steven Goodreau for sharing his code for the Add
Health analysis and for his comments on our work, and we thank the
University of Washington Social Network Modeling Group (Co-PIs: Martina
Morris and Steven Goodreau) for their feedback on this
work. We thank Niel Hens and the POLYMOD project
for sharing the Belgian \mbox{POLYMOD} data. This research uses data
from Add Health, a program project directed by Kathleen Mullan Harris
and designed by J. Richard Udry, Peter S. Bearman and Kathleen Mullan
Harris at the University of North Carolina at Chapel Hill, and funded
by grant P01-HD31921 from the Eunice Kennedy Shriver National Institute
of Child Health and Human Development, with cooperative funding from~23
other federal agencies and foundations. Special acknowledgment is due
Ronald R. Rindfuss and Barbara Entwisle for assistance in the original
design. Information on how to obtain the Add Health data files is
available on the Add Health website
(\url{http://www.cpc.unc.edu/addhealth}). No direct support was
received from grant P01-HD31921 for this analysis.

\begin{supplement}%[id=suppA]
\sname{Supplement A}
\stitle{Model validation and descriptive analyses of simulated contact networks}
\slink[doi]{10.1214/11-AOAS505SUPPA} %[doi,text={...}] - jei reikia
%suskaldyti doi
\slink[url]{http://lib.stat.cmu.edu/aoas/505/supplementA.pdf}
\sdatatype{.pdf}
\sdescription{We compare our fitted degree
distribution to that from an alternate data source, the POLYMOD study.
We compare marginal and joint distributions of variables from contact
networks simulated from our model to the empirical marginal and joint
distributions in the epidemic survey, which was used to estimate model
input parameters.}
\end{supplement}

\begin{supplement}%[id=suppB]
\sname{Supplement B}
\stitle{Sensitivity analysis for targeted antiviral prophylaxis intervention}
\slink[doi]{10.1214/11-AOAS505SUPPB} %[doi,text={...}] - jei reikia
%suskaldyti doi
\slink[url]{http://lib.stat.cmu.edu/aoas/505/supplementB.pdf}
\sdatatype{.pdf}
\sdescription{We perform sensitivity analysis to
assess the impact of the assumption of perfect reporting of contacts in
the targeted antiviral prophylaxis intervention. Simulations are
performed with 90\% and 75\% of contacts reported.}
\end{supplement}

% imsref loaded by lrinkeviciute, 2011-10-05 08:57:48
% imsref loaded by lrinkeviciute, 2011-10-05 10:23:59
%

\printaddresses

\end{document}